\documentstyle[11pt,equation,epsf]{article}
\input{psfig.sty}
\begin{document}
\newcommand{\sx}{\sigma}
\newcommand{\sxa}{\sigma_1}
\newcommand{\sxb}{\sigma_2}
\newcommand{\pha}{\phi_1}
\newcommand{\phb}{\phi_2}
\newcommand{\psa}{\psi_1}
\newcommand{\psb}{\psi_2}
\newcommand{\Psib}{\bar{\Psi}}
\newcommand{\Phib}{\bar{\Phi}}
\newcommand{\mpl}{m_{Pl}}
\newcommand{\Mpl}{M_{Pl}}
\newcommand{\lx}{\lambda}
\newcommand{\ex}{\epsilon}
\newcommand{\be}{\begin{equation}}
\newcommand{\ee}{\end{equation}}
\newcommand{\een}{\end{subequations}}
\newcommand{\ben}{\begin{subequations}}
\newcommand{\beq}{\begin{eqalignno}}
\newcommand{\eeq}{\end{eqalignno}}
\def \lta {\mathrel{\vcenter
     {\hbox{$<$}\nointerlineskip\hbox{$\sim$}}}}
\def \gta {\mathrel{\vcenter
     {\hbox{$>$}\nointerlineskip\hbox{$\sim$}}}}
\pagestyle{empty}
\noindent
\begin{flushright}
CERN--TH/97--301
\\
hep--ph/9710526
\end{flushright} 
\vspace{3cm}
\begin{center}
{ \Large \bf
Two-Stage Inflation 
as a Solution to the \\
\vspace{0.2cm}
Initial Condition Problem  
of Hybrid Inflation
} 
\\ \vspace{1cm}
{\large 
C. Panagiotakopoulos$^{(a)}$ and N. Tetradis$^{(b)}$ 
}
\\
\vspace{1cm}
{\it
(a) Physics Division, School of Technology, University of
Thessaloniki,\\ Thessaloniki 540 06, Greece \\
(b) CERN, Theory Division, 
CH--1211, Geneva 23, Switzerland 
}
\\
\vspace{2cm}
\abstract{
We address the issue of fine-tuning of the initial 
field configuration that can lead to hybrid inflation
in the context of global supersymmetry.
This problem is generated by 
the difference between the energy scale at which the Universe 
emerges from the Planck era 
and the inflationary scale implied by the
COBE observations: $V^{1/4} \sim 10^{-3}\mpl$.
We propose a scenario
with two stages of inflation. The first stage, with a typical scale
not far from $\mpl$, occurs ``naturally'' and provides the 
necessary homogeneity 
for the second stage. The latter generates the 
density perturbations 
that result in the cosmic microwave background
anisotropy observed by COBE.
\\
\vspace{1cm}
}
\end{center}
\vspace{4cm}
\noindent
CERN--TH/97--301 \\
October 1997

\newpage

\pagestyle{plain}
\setcounter{page}{1}

\setcounter{equation}{0}
\renewcommand{\theequation}{{\bf 1.}\arabic{equation}}

\section{Introduction}

The construction of a realistic inflationary scenario 
is constrained by the following main requirements:\\
a) A realistic particle physics model must be constructed, which 
guarantees the presence of a phase where the stress-energy 
tensor of the Universe is dominated by the potential energy density. \\
b) This phase must emerge ``naturally'', so that the
initial conditions for the onset of inflation are not 
fine-tuned. \\
c) The experimental constraints, resulting mainly from the
COBE observation of the cosmic microwave background
anisotropy, must be satisfied. 

The most common way of satisfying the first requirement is
by constructing a field-theoretical model that predicts
an almost flat direction in field space
with non-zero potential energy density. 
A very small slope of the potential
along this direction
results in the slow rolling of the inflaton field $\sigma$.
Inflation stops when the inflaton rolls beyond the point where
the flat direction ends. It is crucial that the flatness of the 
potential in the inflationary trajectory be preserved by 
quantum (and possible thermal) corrections. 
A natural implementation of this scenario 
occurs when hybrid inflation 
\cite{hybrid}
is embedded in the context of supersymmetry
\cite{cop}--\cite{gia}.
Flat directions in the potential appear without fine-tuning, and
are stable against quantum corrections. 
An important aspect of the hybrid inflationary scenario is that the 
part of inflation with observable consequences takes place for 
values of the inflaton field below the Planck scale \cite{cop}
where models can be reliably constructed. 
(Throughout the paper we use the ``reduced'' Planck
scale 
$\mpl= \Mpl/\sqrt{8 \pi}$, $\Mpl = 1.2 \times 10^{19}$ GeV.)

A crucial aspect of the problem of 
initial conditions is whether the presence of inhomogeneities or 
non-zero time derivatives of the fields prevents the onset of
inflation. Previous studies focused on the evolution
of spatial or time derivatives of the inflaton field (for a review
see ref. \cite{gold}). For the ``new'' inflationary scenario
\cite{new,star} the influence of such derivatives makes the
fine-tuning of the initial conditions 
necessary for a sufficient amount of inflation to 
take place \cite{newic,gold}. 
For chaotic inflation \cite{chaotic}, this problem is not severe
\cite{gold}. 
In the context of hybrid inflation, the question of initial conditions 
has been addressed recently \cite{first1}--\cite{nikos}. In 
ref. \cite{nikos} it was shown that 
severe fine-tuning of the initial configuration that will lead to
inflation is necessary. This is a consequence of the 
presence of (one or more) scalar fields coupled to the inflaton,  
and the need to satisfy the observational constraints. 
We briefly summarize the arguments below.

For hybrid inflation 
the slow rolling of the inflaton occurs along a valley of the
potential. It ends when the valley turns into a ridge and 
the slow-roll regime terminates, owing to the growth of
fluctuations of fields orthogonal to 
the inflaton field. Typically this point of instability
corresponds to a value of the inflaton below the 
Planck scale. The 
COBE observation of the cosmic microwave background
anisotropy constrains the properties of the model along the 
inflationary trajectory \cite{star,denpert,report}.
On general grounds, one expects the inflationary energy scale
$V^{1/4}$ 
(determined by the vacuum energy density during inflation) 
to be at least two or three orders of magnitude smaller
than the Planck scale \cite{lyth}:
\be
V^{1/4} / \ex^{1/4} \simeq 7 \times 10^{16}~{\rm GeV}.
\label{oneone} \ee
Here $\ex$ is a ``slow-roll'' parameter \cite{report} that 
must be much smaller than 1 during inflation.
The onset of inflation requires a region of
space of a size of a few Hubble lengths
where the fields take almost constant values, so that the
gradient energy density is negligible compared to the potential energy 
density. 
The earliest time at which one could start talking about such 
regions of space 
is when the Planck era (during which
quantum gravitational fluctuations dominate) ends and 
classical general relativity starts becoming 
applicable. The initial energy density is of order 
$\mpl^4$. For a theory with couplings not much smaller that 1, 
the initial field values within each region
are expected to be of order $\mpl$.

Inflation could start at the end of the Planck era, provided that 
the fields take appropriate values. 
However, it is most likely that the 
fields will evolve, from some initial values that do not give rise to
inflation, to different values that do.
The difference between the initial energy scale $\mpl$ of the field 
evolution and the inflationary scale $V^{1/4}$ implies that 
the fields evolve for a long time before settling down along the 
inflationary trajectory. The Hubble parameter $H$ 
sets the scale for the ``friction'' term in the evolution equations, 
which determines how fast the energy
is dissipated through expansion. 
When the energy density drops much below $\mpl^4$, the
smallness of the ``friction'' term results in a very long  
evolution, during which the fields oscillate around zero 
many times. Some of the trajectories eventually 
settle down in the valley of the potential that produces inflation.
However, the sensitivity to the initial conditions is 
high because of the long evolution. A slight variation of
the initial field values separates inflationary trajectories from 
trajectories that lead to the 
minima of the potential, where inflation does not occur.

The implications for the initial configuration that will lead to the
onset of inflation are severe. It was shown in ref. \cite{nikos} that, for
the prototype model of
hybrid inflation \cite{hybrid} with a scale consistent with the 
COBE observations, 
the most favourable area of inflationary initial conditions is a thin
strip of width $10^{-5}\mpl$ around the $\sigma$ axis. 
Throughout a region of space of a size of the order of the  
Hubble length (which initially is 
$\sim \mpl^{-1}$), 
the initial value of the field orthogonal to the
inflaton must be zero with an accuracy $10^{-5}\mpl$. 
This should be compared to 
the natural scale of the initial fluctuations of the
fields, which is of order $\mpl$.
If this 
condition of extreme homogeneity is not satisfied the fields in
different parts of the original
space region will evolve towards very different values.
In one part they may end up in the valley along the $\sigma$ axis,
while in another they may settle at the minima of the potential. 
Before inflation sets in, the 
size of space regions shrinks compared to the
Hubble distance.
As a result, large inhomogeneities are expected at scales smaller than
$\sim H^{-1}$ when the evolution of the fields finally
slows down. These will prevent the onset of inflation. 
The fine-tuning of the initial configuration must be increased by 
several orders of magnitude if the initial time derivatives
of the fields are non-zero.

In ref. \cite{nikos} the evolution of the scale 
factor $R$ relative to the Hubble parameter was also studied, 
starting from an initial value 
$R_0 \sim H^{-1}_0$. At the onset of
inflation, $R$ was found to be 
smaller than $H^{-1}$ typically by a factor 
of order 10--100.
This implies that the initial homogeneous region should extend far 
beyond a few initial Hubble lengths for this region to inflate.

In this paper we suggest a simple resolution of the issue of fine-tuning
described above. 
The origin of the problem lies in the difference between the inflationary
scale of eq. (\ref{oneone}) implied by the COBE observations, and 
the Planck scale.
This difference has two consequences: \\
a) The ``friction'' term,
proportional to $H$, in the evolution equations 
becomes very small when the energy density drops 
far below the Planck scale. As a result, a long evolution takes place before
the system settles down along
a flat direction or at the minima of the potential.
This generates very high sensitivity to the initial conditions
and requires extreme
homogeneity of the initial configuration, as we explained above. \\
b) The reduction of $R$ relative to $H^{-1}$ during the 
pre-inflationary evolution implies that this extreme
homogeneity must be assumed far beyond the initial Hubble length. \\
The first problem can be resolved if the Hubble parameter stays 
large until the fields settle down along the flat direction 
that will eventually produce inflation. The presence of an
additional field sector with initial energy density of order $\mpl^4$ 
can keep $H$ large during this first part of the evolution.
The second problem indicates the necessity of a first stage of
inflation that will produce a high degree of homogeneity far beyond
the Hubble length. 

The above considerations have motivated us to envisage a scenario
with two stages of inflation. The first stage has a typical scale
$\sim \mpl$, occurs ``naturally'', and provides the homogeneity that 
is necessary for the second stage. The latter generates the 
density perturbations 
that result in the cosmic microwave background
anisotropy observed by COBE. 
Similar ideas have been discussed in ref. \cite{lindec} in the
context of chaotic inflation. 
A scenario analogous to ours 
has been proposed for the resolution of the problem of 
initial conditions for ``new'' inflation \cite{izawa}.

In section 2 we present a quantitative discussion of the fine-tuning
of the initial conditions for hybrid inflation 
in the context of a globally supersymmetric model.
In section 3 we introduce an
extension of the model that allows for an
additional first stage of inflation. This resolves the problem of
fine-tuning. We conclude by discussing the possible implementation of
the idea in the context of supergravity.

\setcounter{equation}{0}
\renewcommand{\theequation}{{\bf 2.}\arabic{equation}}

\section{One-stage inflation and fine-tuning}

We consider a supersymmetric model described by the 
superpotential \cite{cop,shafi}
\be
W = S \left( -\mu^2 + \lx \Phib \Phi \right).
\label{twoone} \ee
Here $S$ , $\Phi$ and $\Phib$ 
are chiral superfields, for which we 
assume canonical kinetic terms. 
The above superpotential is the only renormalizable
one consistent with a continuous $U(1)$ $R$-symmetry 
under which $W \rightarrow e^{i\theta} W$, $S \rightarrow e^{i\theta} S$,
$\Phib \Phi \rightarrow \Phib \Phi$. 
The superfields $\Phi$ and $\Phib$ transform under an internal gauge symmetry,
which we take to be a 
$U(1)$ symmetry for simplicity:
$\Phi \rightarrow e^{i \omega} \Phi$, 
$\Phib \rightarrow e^{-i \omega} \Phib$. The  $S$ superfield is a 
gauge singlet. 
The gauge symmetry is
spontaneously broken
at the scale $\mu/\sqrt{\lx}$. The potential is given by 
\footnote{We use the same notation for superfields and their 
scalar components. Also we have interchanged the notation for
the inflaton and its orthogonal field with respect to ref. \cite{nikos}.}  
\be
V = \left| -\mu^2 + \lx \Phib \Phi \right|^2 + 
\lx^2 \left| S \right|^2
\left( \left| \Phi \right|^2 + \left| \Phib \right|^2 \right)
+ D{\rm-terms}.
\label{twotwo} \ee
Vanishing of the $D$-terms is achieved along the $D$-flat directions 
where $\left| \Phi \right| = \left| \Phib \right|$. 

Through appropriate gauge and $R$-transformations 
along the $D$-flat directions
we can 
bring the fields in the form
\be
S = \frac{\sigma}{\sqrt{2}},
~~~~~~~~~~~~
\Phi = \Phib = \frac{\pha+i\phb}{2}.
\label{twothree}
\ee
The real fields $\sx$, $\pha$, $\phb$ have canonically
normalized kinetic terms. 
The potential can now be written as
\be
V(\sx,\pha,\phb)= \mu^4 - \frac{\lx}{2}\mu^2 \left(\pha^2-\phb^2 \right)
+ \frac{\lx^2}{16} \left(\pha^2+\phb^2 \right)^2
+ \frac{\lx^2}{4}\sx^2 \left(\pha^2+\phb^2 \right).
\label{twofour} \ee
The vacua are located at
$\sx=0$, $\pha^2= 4\mu^2/\lx$, $\phb=0$. 
The potential has a flat direction 
along the $\sx$ axis with
$V(\sx,\pha=\phb=0)=\mu^4$. 
Along this direction the mass terms of the 
$\phi_{1,2}$ fields are
\be
\left[M^2_{\phi}\right]_{1,2}=\mp\lx \mu^2 + \frac{\lx^2}{2} \sx^2.
\label{twofive} \ee
An instability appears for 
\be
\sx^2 < \sx^2_{ins} = \frac{2 \mu^2}{\lx},
\label{twosix} \ee
which can trigger the growth of the $\phi_1$ field.
 
The flatness of the potential along the $\sx$ axis is lifted by 
radiative corrections, resulting from the breaking of supersymmetry 
by the non-zero value of $V(\sx,\pha=\phb=0)$. For a $U(1)$ gauge symmetry
and for $\sx \gg \sx_{ins}$ 
the one-loop contribution to the effective potential along the $\sx$ axis
is given by \cite{shafi,first2}
\be
\Delta V(\sx) = \frac{\lx^2}{16 \pi^2} \mu^4 
\left[ \ln\left(\frac{\lx^2 \sx^2}{2 \Lambda^2} \right) + \frac{3}{2}
\right].
\label{twoseven} \ee
The precise value of the normalization scale $\Lambda$ 
is not important for our discussion. 
The importance of the above contribution for our discussion
lies in that it generates a small positive 
slope along the $\sx$ axis
\be
\Delta V'(\sx)
=\frac{\lx^2}{8 \pi^2} \frac{\mu^4}{\sx}. 
\label{extra} \ee

We assume that a Robertson-Walker metric is a good approximation 
for the regions of space with uniform fields that we are considering. 
The evolution of the fields is given by the standard equations
\beq
\ddot{\sx} + 3 H \dot{\sx} = &~ 
- \frac{\partial V(\sx,\pha,\phb)}{\partial\sx}
\label{twoeight} \\
\ddot{\phi}_{1,2} + 3 H \dot{\phi}_{1,2} = &~ 
- \frac{\partial V(\sx,\pha,\phb)}{\partial\phi_{1,2}}, 
\label{twonine} \eeq
where 
\be
H^2 = \left( \frac{\dot{R}}{R} \right)^2 = ~\frac{1}{3 \mpl^2} \left[
\frac{1}{2} \dot{\sx}^2 
+\frac{1}{2} \dot{\pha}^2
+\frac{1}{2} \dot{\phb}^2
+ V(\sx,\pha,\phb) \right]. 
\label{twoten} \ee

Inflation can occur in the almost flat direction 
along the $\sx$ axis. For $\sx \gg \sx_{ins}$
\footnote{
Inflation can persist down to $\sx_{ins}$ for sufficiently small
coupling $\lx$. For a discussion see refs. \cite{shafi,first2}.
} the ``slow-roll'' parameters 
$\epsilon$, $\eta$ \cite{report} are small as long as 
$\sx/\mpl \gta \lx/\sqrt{8 \pi^2}$.
For $\lx \gta 10^{-3}$
the spectrum of the adiabatic density perturbations 
in this inflationary scenario is estimated to be \cite{shafi}
\be
\delta_H \simeq \sqrt{\frac{N_Q}{75}} 
\frac{4}{\lx}
\left(\frac{\mu}{\mpl}\right)^2,
\label{twoeleven} \ee
where $N_Q\simeq 60$ is the number of e-foldings of our present 
horizon scale during inflation. 
Comparison with the value $\delta_H=1.94 \times 10^{-5}$, 
deduced from the COBE observation of the cosmic microwave background
anisotropy 
gives 
\be
\frac{\mu}{\mpl} \simeq  2.3 \times 10^{-3} \sqrt{\lx}.
\label{twotwelve} \ee

As we have discussed in the introduction, 
inflation does not start immediately, as soon as the 
Universe emerges from the Planck era. An initial evolution
of the fields takes place, during which they approach the
flat direction and eventually settle on the slow-roll trajectory.
A typical example of this part of the evolution is depicted in
fig. 1. 
We have taken $\lx=0.05$ 
and chosen the mass scale $\mu/\mpl=5 \times 10^{-4}$
near the value 
implied by the COBE observations. 
The initial field values are near $\mpl$, with 
$\pha$, $\phb$ smaller than $\sx$, so that the evolution starts
near the flat direction
($\sx_0/\mpl=0.7$, 
$\left[ \pha \right]_0/\mpl=0.1$, 
$\left[ \phb \right]_0/\mpl=0.05$).
The evolution equations (\ref{twoeight})--(\ref{twoten})
have been integrated numerically. The small radiative
contribution of eq. (\ref{twoseven}) has been neglected, so that
$\pha=\phb=0$ corresponds to a completely flat direction. The initial
time derivatives of the fields have been set to zero. This is the 
most favourable situation for the onset of inflation.
We observe an initial stage during which the $\sx$ field decays, while
$\pha$ and $\phb$ oscillate rapidly around zero. The decay is due to
the large initial value of $H$, induced by the large average value
of $V$ in eq. (\ref{twoten}).
When $\sx$ becomes much smaller than its initial value, $H$ 
drops significantly and the ``friction'' term in eqs. (\ref{twoeight}),
(\ref{twonine}) becomes less important. As a result, the system of fields
enters a stage of oscillations around zero, with very slowly
decaying amplitudes. This stage lasts for a 
time larger than the interval depicted in fig. 1 by several orders of
magnitude. Eventually the system
settles down either along the flat direction or at the minima of the
potential. 

This type of evolution generates high 
sensitivity to the initial conditions. 
A slight variation of
the initial field values separates inflationary trajectories from
trajectories that lead to the 
minima of the potential, where inflation does not occur.
This high sensitivity has severe 
implications for the initial configuration that will lead to the
onset of inflation. 
For inflation to start, a region of
space of a size of a few Hubble lengths is required, in which
the fields take almost constant values, so that the
gradient energy density is negligible compared to the potential energy 
density. 
However, it is not 
sufficient to assume that such a homogeneous region emerges at the
end of the Planck era, since this  
homogeneity must be maintained during the
whole evolution, up to the point where inflation starts.
For this to happen, the fields within the region of space should not 
vary more
than the minimal difference between the initial values for a
trajectory that eventually leads to inflation and those for a trajectory
that leads to
the minima of the potential.  
If this
condition of extreme homogeneity is not satisfied, the fields in
different parts of the original
space region will evolve towards very different values.
Before inflation sets in, the 
size of space regions shrinks with respect to the
Hubble distance.
As a result, large inhomogeneities will appear at scales smaller than
$\sim H^{-1}$ when the evolution of the fields finally
slows down, which will prevent the onset of inflation. 

It was shown in ref. \cite{nikos} that, for
the prototype model of
hybrid inflation \cite{hybrid} and $\mu \lta 10^{-1} \mpl$, 
the most favourable area of inflationary initial conditions is a thin
strip around the $\sigma$ axis. If the fields start in this area without
initial time derivatives, $\sx$ does not oscillate around zero, but
quickly settles along the flat direction.
We can obtain a rough estimate of the width of this area if we consider 
eq. (\ref{twoeight}),  and replace $\pha^2$ and $\phb^2$
by their
average values $\left[ \phi_{1,2}^2 \right]_{rms} 
\sim \left[ \phi_{1,2} \right]^2_0/2$ during the evolution.
The equation now reads
\be
\ddot{\sx} + 3 H \dot{\sx} =  
- \frac{\lx^2}{4} 
\left( \left[ \pha \right]^2_0 + \left[ \phb \right]^2_0
\right) \sx.
\label{twothirteen} \ee
Two time scales
characterize the solutions of this equation.
The first one is related to the friction term and is given by  
$t_H^{-1} = 3H/2$. For $\left[ \phi_{1,2} \right]_0 \ll 2 \mu/\sqrt{\lx}$ 
and 
$\lx^2 \sx^2 \left( \left[ \pha \right]^2_0 + \left[ \phb \right]^2_0
\right)/8 \ll \mu^4$,
we have 
\be
t_H =  \frac{2}{\sqrt{3}}  \frac{\mpl}{\mu^2}.
\label{twofourteen} \ee
The other time scale is obtained if we neglect the friction term and 
consider the oscillations of the $\sx$ field.
One-fourth of the period is the typical time for the system to roll
to the origin and away from an inflationary solution. It is given
by 
\be 
t_{osc} = \frac{\pi}{\lx} 
\frac{1}{\sqrt{\left[ \pha \right]^2_0 + \left[ \phb \right]^2_0}}.
\label{twofifteen} \ee
Inflation sets in if $t_{osc} \gta t_H$, which gives 
\be
\frac{\sqrt{\left[ \pha \right]^2_0 + \left[ \phb \right]^2_0}}{\mpl} 
\lta \frac{\sqrt{3}~\pi}{2~\lx} 
\left( \frac{\mu}{\mpl} \right)^2.
\label{twosixteen} \ee
We have verified numerically that the above relation gives the correct 
order of magnitude for the size of the strip around the $\sx$ axis that
leads to inflationary solutions. 
Our assumptions for the derivation of the above bound break 
down when 
$\sx \gta \sqrt{24/\pi^2}  \mpl = 1.6\mpl$.

For our choice of $\mu$ and $\lx$ 
the initial value of the fields orthogonal to the
inflaton must be zero with an accuracy $\sim 10^{-5}\mpl$ 
throughout a region of space of a size of the order of the  
Hubble length.
This should be compared to 
the natural scale of the initial fluctuations of the
fields, which is of order $\mpl$.
The fine-tuning of the initial configuration must be increased by 
several orders of magnitude if the initial time derivatives
of the fields are non-zero \cite{nikos}.
In ref. \cite{nikos} the evolution of the scale 
factor $R$ relative to the Hubble parameter was also studied, 
starting from an initial value 
$R_0 \sim H^{-1}_0$. At the onset of
inflation, $R^{-1}$ was found to be 
smaller than $H^{-1}$ typically by a factor 
of order 10--100.
This implies that the initial homogeneous region should extend far 
beyond a few initial Hubble lengths for this region to inflate.
It is very difficult
to calculate a probability distribution for the initial 
field configurations, and thus have a 
quantitative estimate of the probability of inflation to occur. 
This requires a reliable description of Planck scale dynamics. 
For this reason we cannot address the question of whether such a homogeneous
initial state is probable or not.
The above results, however, indicate that one-stage hybrid inflation 
does not occur ``naturally''.
In the following section we describe two possible 
extensions of the model,
within which 
the problem of the fine-tuning of the initial conditions is resolved.

\setcounter{equation}{0}
\renewcommand{\theequation}{{\bf 3.}\arabic{equation}}

\section{Two-stage inflation}
We consider two extensions of the model of the previous section, which
permit a two-stage inflation. 

\subsection{Chaotic-Hybrid}
The first model is described by the superpotential 
\be
W = S \left( -\mu^2 + \lx \Phib \Phi \right) + \frac{1}{2} m S'^2,
\label{threeone} \ee
where $S'$ is a gauge singlet chiral superfield, whose
scalar components are given by 
$S'=\left( \sxa + i\sxb \right)/\sqrt{2}$.
This is a very simple extension of the minimal superpotential
giving rise to hybrid inflation.  
The above superpotential is the only renormalizable
one consistent with a continuous $U(1)$ $R$-symmetry 
under which $W \rightarrow e^{i\theta} W$, $S \rightarrow e^{i\theta} S$,
$S' \rightarrow e^{i\theta/2} S'$,
$\Phib \Phi \rightarrow \Phib \Phi$. 
The potential derived from $W$ is given by eq. (\ref{twofour})
with the addition of the mass term
\be
\Delta V(\sxa, \sxb) = \frac{1}{2} m^2 \left( \sxa^2 + \sxb^2 \right).
\label{threetwo} \ee

For initial values of $\sxa$ and $\sxb$
larger than $\mpl$, this system undergoes a first stage of 
chaotic inflation, followed by a second stage of 
hybrid inflation similar to the one discussed in the
previous section. In fig. 2 we present the evolution of the various 
fields for the choice of parameters
$\mu/\mpl=5\times 10^{-4}$, 
$\lx=0.05$, $m/\mpl=10^{-3}$.
The initial values of the fields have been taken
to be $\sx_0/\mpl=0.7$, 
$\left[ \pha \right]_0/\mpl=0.1$, 
$\left[ \phb \right]_0/\mpl=0.05$,
$\left[ \sxa \right]_0/\mpl=6$, 
$\left[ \sxb \right]_0/\mpl=5$, with
their time derivatives equal to zero. 
We observe that the fields $\sxa$, $\sxb$ quickly 
approach a slow-roll regime and they attain almost constant 
velocities. This regime terminates when $\sx_{1,2} \simeq 1$. 
During this time, the vacuum energy density is dominated by the
contribution of eq. (\ref{threetwo}) and a period of
inflation takes place. The Hubble parameter
has a value $H_i/\mpl \simeq 3.4 \times 10^{-3}$ at the beginning of the
slow-roll regime, which drops to $H_f/\mpl \simeq 6 \times 10^{-4}$ 
at its end. A total number of $N_1\simeq 16$ e-foldings is generated by this
first stage of inflation. 

In the model we have described, the onset of inflation occurs ``naturally''.
The initial correlation lengths of the various fields
are comparable to the Hubble length $H^{-1}$ 
and, therefore,
an initial field configuration homogeneous over a few Hubble lengths is
not unlikely. 

The evolution of $\sx$, $\pha$, $\phb$ is also
depicted in fig. 2. 
The presence of a large ``friction'' term 
$H \sim m \sqrt{\sxa^2+\sxb^2}/\mpl$
in their evolution equations
forces these fields to quickly settle down along their flat direction. 
The time derivatives of $\pha$, $\phb$ go to zero, while that of
$\sx$ reaches a very small value
because of the small slope induced
by the contribution of eq. (\ref{extra}). The total classical
evolution of $\sx$ during the first stage of inflation is
negligible.  
This behaviour should be contrasted with the one depicted in fig. 1.
In that case the ``friction'' term $H \sim \mu^2/\mpl$
was smaller by several orders of magnitude, and the field oscillations
decayed very slowly. 

After the end of the first stage of inflation, an intermediate stage
takes place during which 
$\sxa$ and $\sxb$ oscillate around zero. 
The energy density of
the Universe decreases, until the point where the final vacuum 
energy density $\mu^4$ starts to dominate and the second stage of
inflation begins.
During all this time the system $(\sx,\phi_{1,2})$ remains unperturbed
at the classical level, as it is not coupled to $\sxa$, $\sxb$. 
However, a more careful analysis is required in order to take into
account quantum fluctuations of the fields. 

During the first stage of inflation, the almost massless field 
$\sx$ has a spectrum of quantum-mechanical fluctuations
characterized by 
\be
\left( \Delta \sx \right)^2_k = \left( \frac{H_1}{2 \pi} \right)^2,
\label{threethree} \ee
where $H_1$ is the Hubble parameter. 
For constant $H_1$, 
the freezing of fluctuations that cross the horizon results in  
a mean square fluctuation of the classical field $\sx$ \cite{freeze}
\be
\left( \Delta \sx \right)^2 = N_1 \left(\frac{H_1}{2 \pi} \right)^2,
\label{threefour} \ee
where $N_1$ is the number of e-foldings.
For the model we are discussing $H_1$ is slowly decreasing, and the above
expression gives an upper bound on $\Delta \sx$ if the initial value
$H_i$ is employed. For $N_1 \simeq 16$, $H_1=H_i
 \simeq 3.4 \times 10^{-3} \mpl$
we conclude that the mean square fluctuation of $\sx$ is negligible
compared to its mean value $\sx\simeq 0.63$. 
At the end of the first stage of inflation, 
the energy density in spatial gradient terms 
associated with the perturbations $\sim H_f/2\pi$
of massless fields like $\sx$ is $\sim H^4_f/4\pi^2$. 
It is further reduced by the subsequent expansion and
stays much smaller than the total energy density during the 
whole intermediate stage between the two inflations.  
As a result, it is negligible compared to the vacuum energy density 
$\mu^4$ when this begins to dominate.

The fluctuations of the massive fields $\phi_i,~i=1,2$, generated by the
first stage of inflation are given by 
\be
\left( \Delta \phi_i \right)^2_k = \left( c~\frac{H_1^2}{m_i} \right)^2, 
\label{threefive} \ee
where $m_i$ is their mass and $c={\cal O} \left( 10^{-1} \right)$.
During the intermediate stage the amplitude of these
fluctuations drops $\sim R^{-3/2}$. As a result we expect that, when
the energy density becomes comparable to 
$\mu^4$, this amplitude is approximately given by 
\be
\left( \Delta \phi_i \right)_k
\simeq c~\frac{H^2_1}{m_i} \left( \frac{H_2}{H_1} \right)^{\frac{1}{1+w}}.
\label{threesix} \ee
Here $H_1$ has to be taken close to its value towards the
end of the first stage of
inflation $H_1 \simeq H_f$. 
$H_2$ is the Hubble parameter during the second
stage, and $w$ is determined by the dynamics of the intermediate stage.
For a system of massive oscillating fields, or a matter-dominated
Universe, $w=0$. For a radiation-dominated Universe, $w=1/3$.
For the model we are considering $H_2/\mpl=1.4\times 10^{-7}$
and the fluctuations of $\pha$, $\phb$
are much smaller than the bound of eq. (\ref{twosixteen}). 

We conclude that the necessary conditions for the onset of the
second stage of inflation are ``naturally'' satisfied.
Due to the expansion during the first stage, the homogeneous regions
extend far beyond the Hubble length $H^{-1}_2$. The fields $\sx$, 
$\pha$, $\phb$ are localized on the flat direction of the potential. 
More specifically, the degree of homogeneity of  
$\pha$, $\phb$ satisfies the constraint of eq. (\ref{twosixteen}). 
Thus, the second stage of inflation ``naturally'' sets in. 
It generates
$N_2 \simeq 6.3 \times 10^3$ 
e-foldings and density perturbations in agreement with the COBE
observations.

\subsection{Hybrid-Hybrid}
The model described by the
superpotential 
\be
W = S \left( -\mu^2 + \lx \Phib \Phi \right)
+S' \left( -\mu'^2 + \lx' \Psib \Psi + g \Phib \Phi \right)
\label{threeseven} \ee
is essentially composed of two sectors 
similar to the one of eq. (\ref{twoone}). 
We assume that the new superfields $\Psi$, $\Psib$ 
transform under an internal $U(1)$ gauge symmetry. 
The  $S'$ superfield is a 
gauge singlet. 
Under the continuous $U(1)$ $R$-symmetry 
the new superfields transform as $S' \rightarrow e^{i\theta} S'$,
$\Psib \Psi \rightarrow \Psib \Psi$.
The absence of a coupling $S \Psib \Psi$ is explained if we
regard $S$ as the linear combination of the gauge singlets 
that does not couple to $\Psib \Psi$.

Making use of the continuous symmetries of $W$ and staying 
along the $D$-flat directions, we can choose the 
real components of the various canonically normalized
scalar fields as
\beq
S =&~\frac{\sigma}{\sqrt{2}},
~~~~~~~~~~~~~~~~~~~
\Phi = \Phib = ~\frac{\pha+i\phb}{2},
\nonumber \\
S' =&~\frac{\sxa+i\sxb}{\sqrt{2}},
~~~~~~~~~~~~
\Psi = \Psib = ~\frac{\psa+i\psb}{2}.
\label{threeeight} \eeq
The potential is then given by the expression 
\beq 
V=
&~~\mu'^4-\frac{g}{2}\mu'^2 \left(\pha^2-\phb^2 \right)
+\frac{g^2}{16} \left( \pha^2+\phb^2 \right)^2
\nonumber \\
&-\frac{\lx'}{2} \mu'^2 \left(\psa^2-\psb^2 \right) 
+ \frac{\lx'^2}{16} \left(\psa^2+\psb^2 \right)^2
+\frac{\lx' g}{2} \left[
\frac{1}{4} \left(\pha^2-\phb^2 \right) \left(\psa^2-\psb^2 \right) 
+ \pha \phb \psa \psb \right]
\nonumber \\
&+\mu^4-\frac{\lx}{2}\mu^2 \left(\pha^2-\phb^2 \right)
+\frac{\lx^2}{16} \left( \pha^2+\phb^2 \right)^2
\nonumber \\
&+\frac{\lx'^2}{4} \left( \sxa^2+\sxb^2 \right) \left( \psa^2+\psb^2 \right)
+\frac{1}{4} \left[ \left( \lx \sx + g \sxa \right)^2 + g^2 \sxb^2 \right]
\left( \pha^2 +\phb^2 \right),
\label{threenine} \eeq
where the mass scales $\mu'$, $\mu$ are chosen to satisfy the 
inequality  $\mu' \gg \mu$.
The minima of this potential are located at
\beq 
\sx=&0,~~~~~~~~~~~~~~~~~\pha^2= \frac{4}{\lx}\mu^2,~~~~~~~~~
~~~~~~~~~~~~~~~~~~~~~~\phb=0, 
\nonumber \\
\sxa=&\sxb=0,~~~~~~~~~~
\psa^2=\frac{4}{\lx'}\mu'^2
-\frac{4g}{\lx\lx'}\mu^2 \simeq \frac{4}{\lx'}\mu'^2,
~~~~~~~\psb=0.
\label{threeten} \eeq

For $\pha=\phb=\psa=\psb=0$ the potential is independent of 
$\sx$, $\sxa$, $\sxb$ and has the value
$V=\mu'^4+\mu^4 \simeq \mu'^4$, which is the vacuum energy density during
the first stage of inflation. 
The mass terms of the $\phi_{1,2}$ and $\psi_{1,2}$ fields are
\beq
\left[M^2_{\phi}\right]_{1,2}=~&
\mp g \mu'^2 \mp \lx \mu^2
+\frac{1}{2} \left[ \left( \lx \sx + g \sxa \right)^2 + g^2 \sxb^2 \right]
~\simeq~ \mp g \mu'^2
+\frac{1}{2} \left[ \left( \lx \sx + g \sxa \right)^2 + g^2 \sxb^2 \right]
\nonumber \\
\left[M^2_{\psi}\right]_{1,2}=~&
\mp \lx' \mu'^2 
+ \frac{\lx'^2}{2} \left( \sxa^2 + \sxb^2 \right).
\label{threeeleven} \eeq
We see that the squared mass term of the $\psa$ field becomes negative for 
\be
\sxa^2 + \sxb^2 < \frac{2 \mu'^2}{\lx'},
\label{threetwelve} \ee
while that of the $\pha$ field turns negative for
\be
\left( \sxa + \frac{\lx}{g} \sx \right)^2 + \sxb^2 < \frac{2 \mu'^2}{g},
\label{threethirteen} \ee
signalling instabilities in the $\psa$ and $\pha$ directions. 

The flatness of the potential is lifted by radiative corrections.
For $\sx$ and $\sx_{1,2}$ far above the instability points
the one-loop 
contribution to the effective potential is
\beq
\Delta V(\sx,\sxa,\sxb) \simeq &~~~~ 
\frac{g^2}{16 \pi^2}  \mu'^4
\left[ \ln\left(
\frac{(\lx \sx+ g\sxa)^2+g^2\sxb^2}{2 \Lambda^2} 
\right) 
+ \frac{3}{2}
\right]
\nonumber \\
&+\frac{\lx'^2}{16 \pi^2}  \mu'^4 
\left[ \ln\left(
\frac{\lx'^2 (\sxa^2+\sxb^2)}{2 \Lambda^2} 
\right) 
+ \frac{3}{2}
\right].
\label{threefourteen} \eeq

The second stage of 
inflation can take place for
$\pha=\phb=\sxa=\sxb=\psb=0$,  
$\psa^2=4\mu'^2/\lx'$. Then the potential is independent of
$\sx$ and the vacuum energy density is
$V=\mu^4$. A small slope in the $\sx$ direction is
provided by the radiative contribution of eq. (\ref{extra}),
leading to a second stage of inflation in complete analogy to the one 
we discussed in section 2.

In figs. 3 and 4 we present the evolution 
of the various fields for the 
model of eq. (\ref{threeseven}) with
$\mu/\mpl=5\times 10^{-4}$, 
$\lx=0.05$, $\mu'/\mpl=0.1$, $\lx'=0.1$, $g=0.1$,  
and initial conditions
$\sx_0/\mpl=0.7$, 
$\left[ \pha \right]_0/\mpl=0.1$, 
$\left[ \phb \right]_0/\mpl=0.05$,
$\left[ \sxa \right]_0/\mpl=0.7$, 
$\left[ \sxb \right]_0/\mpl=0.6$, 
$\left[ \psa \right]_0/\mpl=0.04$, 
$\left[ \psb \right]_0/\mpl=0.03$. 
The initial time derivatives of the fields have
been taken equal to zero. 
We observe that the system quickly settles along the
flat direction $\pha=\phb=\psa=\psb=0$ and a first stage of
inflation begins. 
As the initial correlation lengths of the various fields
are comparable to the initial Hubble length $H^{-1}$, 
it is not unlikely to have 
a field configuration homogeneous over a few Hubble lengths.
The initial homogeneity 
is expected to be preserved by the short evolution to the flat direction.
As a result, the onset of inflation occurs ``naturally'' in this model.

The first stage of inflation is characterized by a 
constant Hubble parameter $H_1/\mpl=5.8\times 10^{-3}$. It 
lasts until an instability point is
reached. For our choice of couplings the mass term of the 
$\psa$ field is the first one to turn negative. This occurs for
$\sx/\mpl=0.61$, $\sxa/\mpl=0.32$, $\sxb/\mpl=0.31$.
A total number of $N_1\simeq 1.5\times 10^3$ e-foldings is generated by the
first stage of inflation. 
The subsequent evolution proceeds in complete analogy to the discussion
in the previous section. An intermediate stage takes place, during which
the fields $\sx_{1,2}$, $\psi_{1,2}$ oscillate around the minimum at 
$\sxa=\sxb=\psb=0, \psa^2=4\mu'^2/\lx'$, and
the energy density is dissipated through expansion. When it becomes
$\sim \mu^4$ the second stage of inflation begins, with 
a constant Hubble parameter $H_2/\mpl=1.4 \times 10^{-7}$.
Due to the expansion during the first stage, the homogeneous regions
extend far beyond the Hubble length $H^{-1}_2$. Moreover, the fields $\sx$, 
$\pha$, $\phb$ are localized on the flat direction of the potential. 
The amplitudes of 
$\pha$, $\phb$ are given by 
eq. (\ref{threesix}) and satisfy the constraint of eq. (\ref{twosixteen}). 
Thus, the second stage of inflation ``naturally'' sets in. It generates
$N_2 \simeq 5.9 \times 10^3$ 
e-foldings and density perturbations in agreement with the COBE
observations.

\setcounter{equation}{0}
\renewcommand{\theequation}{{\bf 4.}\arabic{equation}}

\section{Conclusions}

In this paper we addressed the problem of 
fine-tuning of the initial conditions for supersymmetric hybrid inflation. 
This problem is generated by 
the difference between the energy scale at which the Universe 
emerges from the Planck era (near $\mpl$) 
and the inflationary scale implied by the
COBE observations ($V^{1/4} \sim 10^{-3}\mpl$).
We suggested a simple resolution of the issue of fine-tuning
by considering a scenario
with two stages of inflation at two different energy scales.
The first stage has a typical scale
not far from $\mpl$. 
Consequently, the Hubble parameter stays 
large until the fields settle down along the direction 
that produces inflation. Due to the large ``friction'' term
in the equations of motion, the initial part of the evolution
towards the inflationary direction 
is short and the first stage of inflation 
occurs ``naturally''. 
After the end of this stage a subset of the fields moves 
towards a minimum of the potential, around which it performs 
damped oscillations.  
During this intermediate 
stage the energy density is reduced 
through expansion. A second stage of inflation 
begins when the energy density
falls below the false vacuum energy density associated with
a second-order phase transition involving the
remaining fields.  
The homogeneity far beyond the Hubble length that was produced
during the first inflationary stage 
makes the onset of the second stage ``natural'', despite the
fact that the Hubble length is much larger than the one of the 
first stage. 
The second stage of inflation generates the density perturbations 
that result in the cosmic microwave background
anisotropy observed by COBE.
In section 3 we gave two realizations of the above scenario
in the context of global supersymmetry. The first employs one chaotic
and one hybrid inflation, while the second one employs two hybrid 
inflations that take place for field values below $\mpl$. 

The generalization of the two-stage inflationary scenario
in the context of supergravity is difficult. The case of
a first-stage chaotic inflation suffers from the obvious difficulty 
that it requires field values larger than $\mpl$.
For the second scenario that 
involves two hybrid inflations, the tree-level potential 
has a multiple $F$-flat direction at $\Phi=\Phib=\Psi=\Psib=0$.
When global $N=1$ supersymmetry is replaced by $N=1$ supergravity,  
all flat directions are, in general, lifted and inflation 
becomes impossible.  In the context of canonical supergravity 
(minimal K\"ahler potential) and a linear superpotential of 
the form $W=-\mu^2 S$ (like the one encountered in hybrid 
inflationary models) a miraculous cancellation takes place that
prevents the appearance of a mass term for $S$ \cite{cop}. Therefore,
there is a possibility for $S$ to play the role of the inflaton. 
However, for a superpotential such as the one of eq. (\ref{threeseven})
only a linear 
combination of $S$ and $S'$ stays ``massless'', with the orthogonal
combination acquiring a large mass term \cite{costas}. This implies
that only one inflationary stage is likely to survive in the context of 
supergravity. However, this argument does not apply to the case where inflation
is driven by a $D$-term
energy density \cite{dterm}. 
The two-stage inflationary scenario 
may then be possible along an appropriate combination of $F$-flat and
$D$-flat directions.

\vspace{0.5cm}
\noindent
{\bf Acknowledgements}: 
We would like to thank G. Dvali, J. Garcia-Bellido
and G. Lazarides for useful discussions. 
This research was supported in part by the E.U.
under TMR contract No. ERBFMRX--CT96--0090.

\newpage

\pagestyle{empty}

\begin{figure}
\psfig{figure=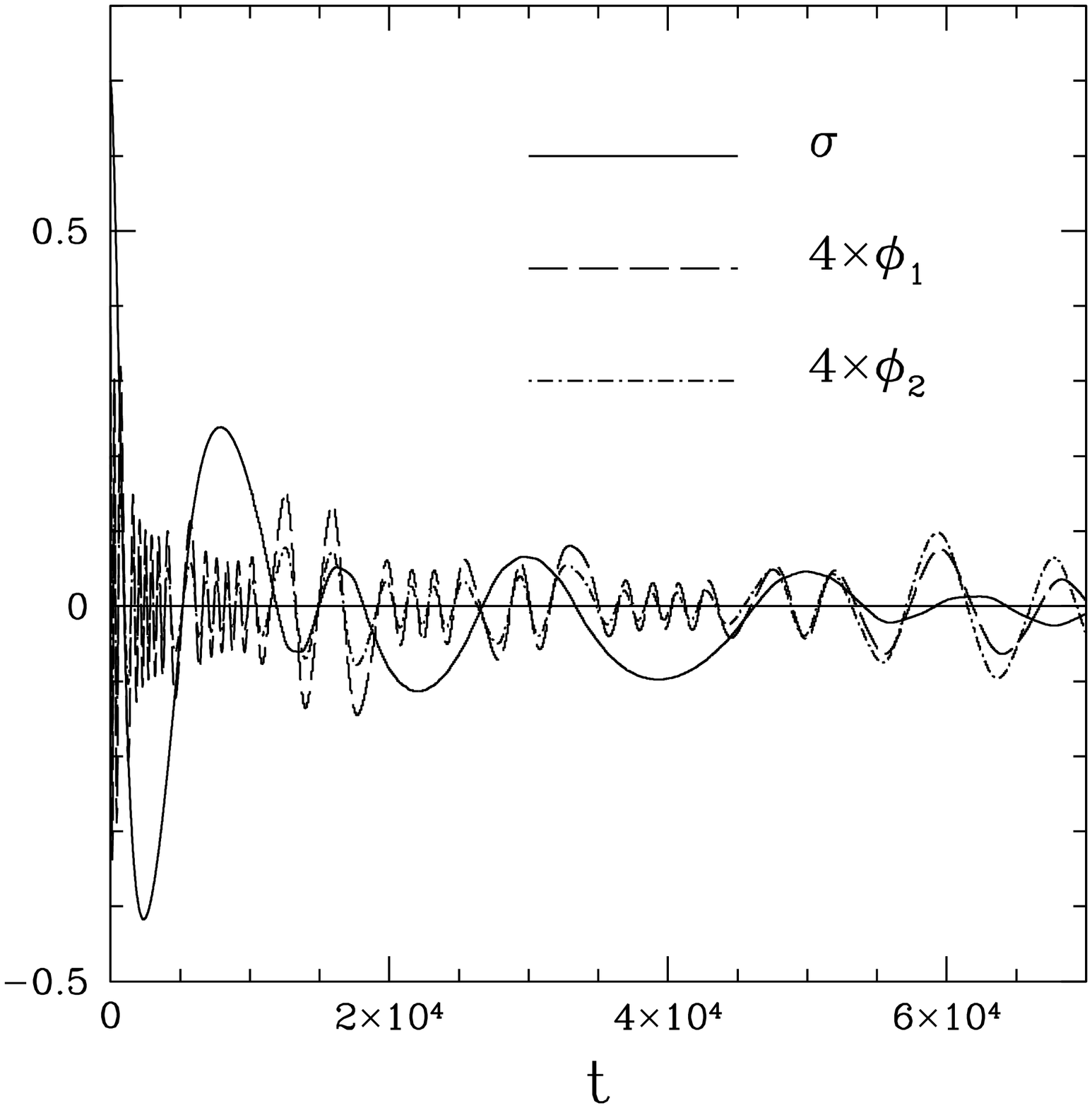,height=16.0cm}
\vspace*{2cm}

Fig. 1: The evolution of $\sx$, $\pha$, $\phb$ (in units of $\mpl$), for the
theory of eq. (\ref{twoone}) with
$\mu/\mpl=5\times 10^{-4}$, 
$\lx=0.05$ 
and initial conditions
$\sx_0/\mpl=0.7$, 
$\left[ \pha \right]_0/\mpl=0.1$, 
$\left[ \phb \right]_0/\mpl=0.05$. 
\end{figure}

\begin{figure}
\psfig{figure=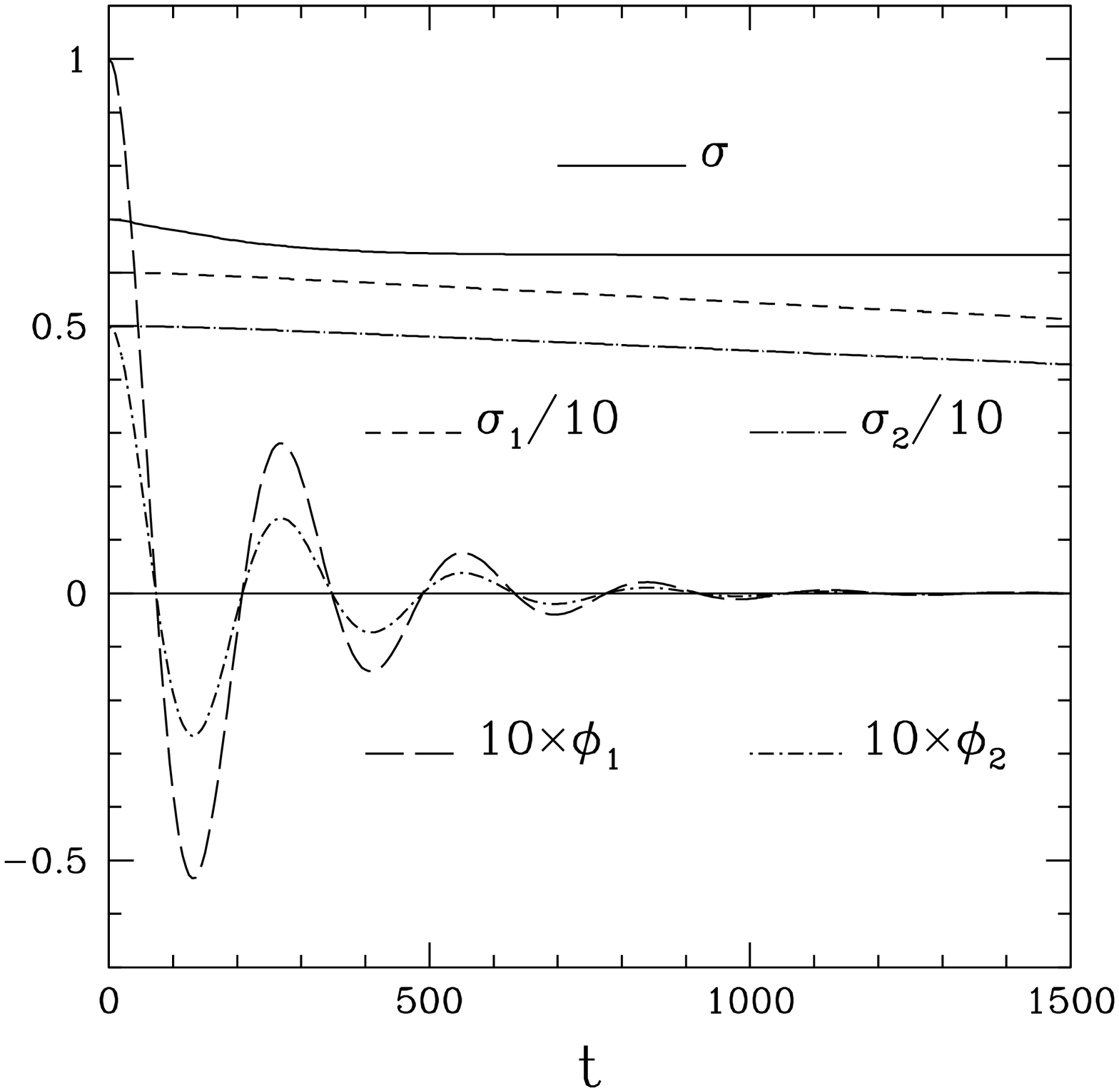,height=16.0cm}
\vspace*{2cm}

Fig. 2: The evolution of $\sx$, 
$\sxa$, $\sxb$, $\pha$, $\phb$ (in units of $\mpl$),
for the
theory of eq. (\ref{threeone}) with
$\mu/\mpl=5\times 10^{-4}$, 
$\lx=0.05$, $m/\mpl=1\times 10^{-3}$,
and initial conditions
$\sx_0/\mpl=0.7$, 
$\left[ \pha \right]_0/\mpl=0.1$, 
$\left[ \phb \right]_0/\mpl=0.05$,
$\left[ \sxa \right]_0/\mpl=6$, 
$\left[ \sxb \right]_0/\mpl=5$. 
\end{figure}

\begin{figure}
\psfig{figure=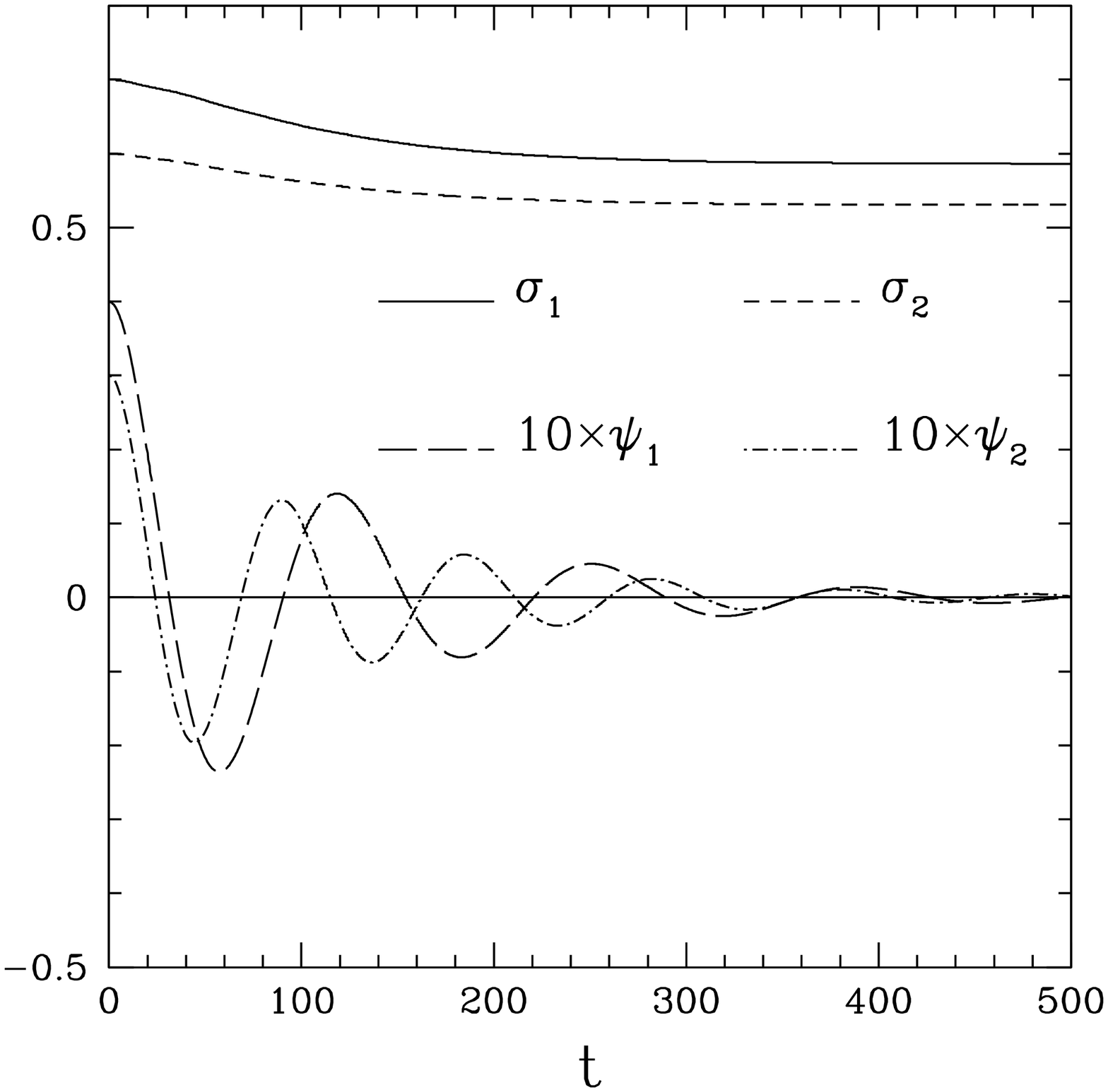,height=16.0cm}
\vspace*{2cm}

Fig. 3: The evolution of $\sxa$, $\sxb$, $\psa$, $\psb$ (in units of $\mpl$),
for the
theory of eq. (\ref{threeseven}) with
$\mu/\mpl=5\times 10^{-4}$, 
$\lx=0.05$, $\mu'/\mpl=0.1$, $\lx'=0.1$, $g=0.1$,  
and initial conditions
$\sx_0/\mpl=0.7$, 
$\left[ \pha \right]_0/\mpl=0.1$, 
$\left[ \phb \right]_0/\mpl=0.05$,
$\left[ \sxa \right]_0/\mpl=0.7$, 
$\left[ \sxb \right]_0/\mpl=0.6$, 
$\left[ \psa \right]_0/\mpl=0.04$, 
$\left[ \psb \right]_0/\mpl=0.03$. 
\end{figure}

\begin{figure}
\psfig{figure=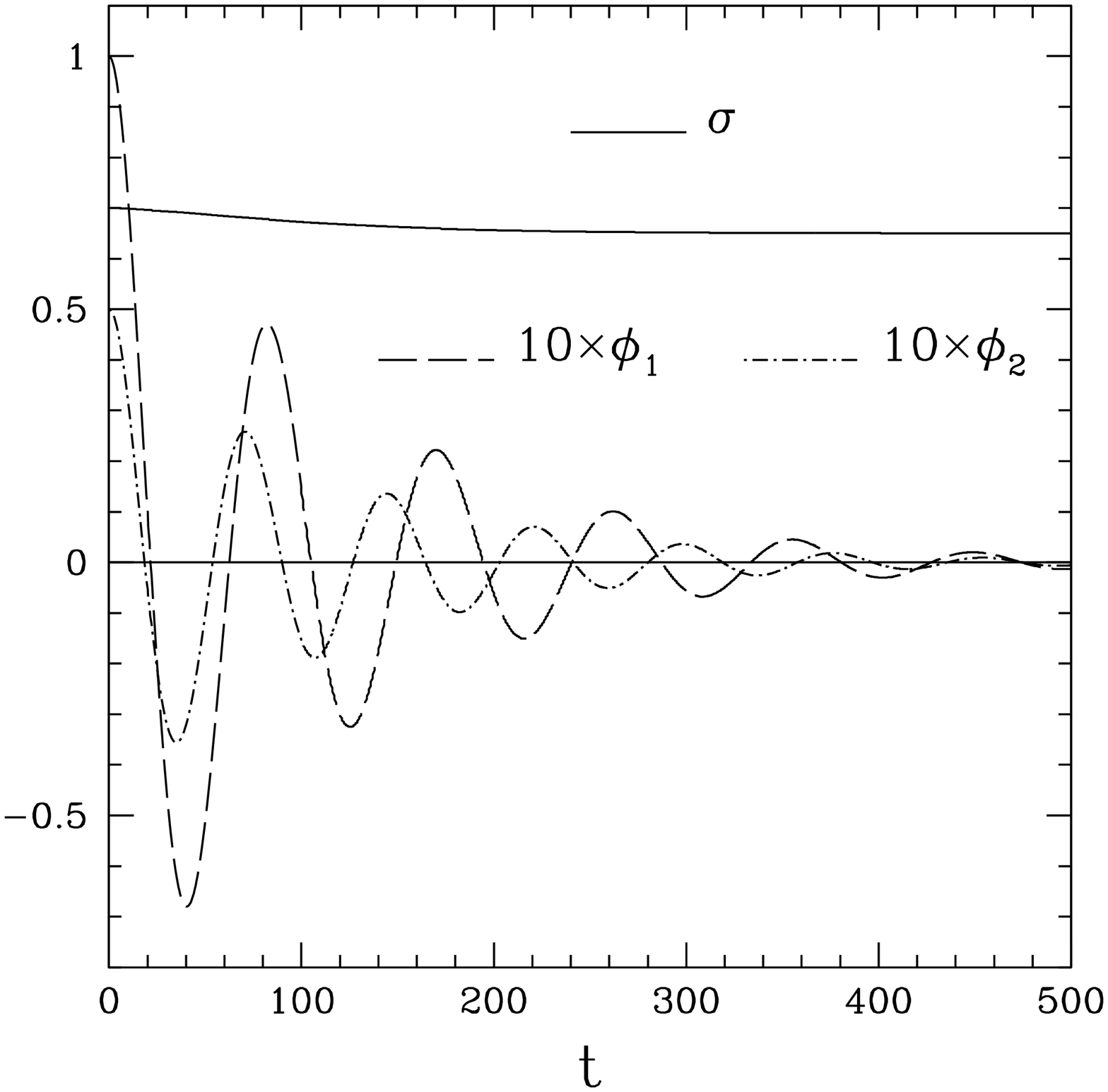,height=16.0cm}
\vspace*{2cm}

Fig. 4: The evolution of $\sx$, $\pha$, $\phb$ (in units of $\mpl$),
for the
theory of fig. 3.
\end{figure}

\end{document}